
\magnification=1200
\overfullrule = 0 pt
\baselineskip=24 true bp
\hfuzz=0.5pt

\def\etal{{\it et al.\/}}
\def\ie{{\it i.e.\/}}
\def\av#1{{\langle{#1}\rangle}}
\def\NI{\noindent}
\long\def\UN#1{$\underline{{\vphantom{\hbox{#1}}}\smash{\hbox{#1}}}$}

\centerline{\bf Two-Species Annihilation with Drift: A Model with Continuous}
\centerline{\bf Concentration-Decay Exponents}
\vskip 1cm
\centerline{Daniel ben-Avraham,\footnote*{e-mail:
{\sl qd00@craft.camp.clarkson.edu}} Vladimir Privman,\footnote{$^{\S}$}{e-mail:
{\sl privman@albert.phy.clarkson.edu}} and Dexin
Zhong\footnote{$^{\ddag}$}{e-mail: {\sl zhongd@craft.camp.clarkson.edu}}}
\bigskip
\centerline{
Physics Department, Clarkson University}
\centerline{
Potsdam, New York 13699--5820, USA}
\vskip 1.2cm
\noindent{\bf ABSTRACT:} We propose a model for diffusion-limited annihilation
of two species, $A+B\to A$ or $B$, where the motion of the
particles is subject to a drift.  For equal initial
concentrations of the two species, the density follows a
power-law decay for large times. However, the decay exponent varies
\UN{continuously} as a
function of the probability of which particle, the hopping one or the target,
survives in the reaction. These results suggest
that diffusion-limited reactions subject to drift do not fall into a limited
number of universality classes.

\bigskip
\NI PACS:\ \ 82.20.Mj,\ \ 05.70.Ln,\ \  02.50.--r,\ \ 68.10.Jy,
\vfill\eject

Diffusion-limited reactions (DLR) in low dimensions have been extensively
studied because of their anomalous kinetics, and because they serve as
simple prototypes of complex, nonequilibrium dynamical systems [1-10].
Recently, attention has focused on  anisotropic DLR: the particles move
preferentially in one spatial direction, giving rise to an effective
drift [11-16].

{}For the one-species DLR of coalescence; $A+A\to A$, and annihilation; $kA\to
0$, it has been shown that the spatial anisotropy introduces no appreciable
changes [11, 15-16]. However, the two-species annihilation process, $A+B\to 0$,
exhibits markedly different behavior with or without
drift [12-14]. In one dimension, when the initial concentrations of $A$ and
$B$ particles are equal, the total concentration decays as $t^{-1/4}$ in the
isotropic case, and as $t^{-1/3}$ in the presence of drift.  This interesting
behavior is attributed largely to the hard core repulsion between particles of
like species.  A heuristic explanation, based on the Burgers equation, for the
$t^{-1/3}$ decay has been presented along these lines [13-14]. A recent exact
analysis of a two-species annihilation model with no hard core interactions
confirms their essential role: in their absence, the concentration decays as
$t^{-1/4}$, with or without the drift [16].

In this work, we introduce a new anisotropic two-species annihilation model,
with
hard core interactions between like species.  A distinct characteristic of our
model is that it contains an adjustable parameter.  We find that the
concentration power-law decay exponent depends \UN{continuously} on this
parameter. Nonuniversal exponents are a surprising feature.  Our finding is
based  on extensive numerical Monte Carlo simulations to be detailed below.

%
Our model is defined on a one-dimensional lattice. Each lattice site can be in
one of the three states: empty (0), occupied by a
single $A$ particle, or occupied by a single $B$
particle. Numerical results were obtained for the case of maximum
anisotropy [12-14], where
particles may hop only to the nearest site to their right:
$$
\cdots A0\cdots\to \cdots 0A\cdots,\qquad \cdots B0\cdots\to \cdots 0B\cdots,
\eqno(1)
$$
with equal rates for $A$ and $B$. Hopping is disallowed if the target site is
occupied by a particle of the same species; this models hard core
interactions. If the target site is occupied by a particle of the opposite
species, hopping is allowed and reaction takes place with outcome
determined by the probabilistic rule:
$$
\cdots AB\cdots\to\cases{\cdots 0A\cdots &prob. $p$,\cr
                         \cdots 0B\cdots &prob. $1-p$,\cr}
\>{\rm and}\>
\cdots BA\cdots\to\cases{\cdots 0B\cdots &prob. $p$,\cr
                         \cdots 0A\cdots &prob. $1-p$.\cr}
\eqno(2)
$$
The reaction may then be represented as
$$
A+B\to A{\rm\ or\ }B,
\eqno(3)
$$
where the product species is that of the hopping particle, with
probability $p$, or that of the target particle, with probability
$1-p$. Thus, $p$ may be thought of as a
parameter which represents the ``persistence'' of the hopping particle.
Stoichiometrically, however, the reaction rule is equivalent to the symmetric
annihilation reaction
$$
A+B\to {1\over 2}A+{1\over 2}B,
\eqno(4)
$$
regardless of the value of $p$, because the number of $AB$ nearest-neighbor
pairs is equal to the number of
$BA$ nearest pairs (assuming homogeneous initial distributions),
even for unequal concentrations of $A$ and $B$ particles.

%
We have performed concurrent simulations of the above model on a cluster of
over 50 IBM
RS6000 workstations, for lattices of $10^6$ sites, with
periodic boundary conditions.  Simulations
ran up to times $t=10^6$. Particle hopping was random and independent,
with the rate defined so that in one time unit each of the surviving
particles performs an average of one hopping attempt. The attempt could
lead to motion, reaction, or be discarded, depending on the state of the target
site. Data from about 50 such runs were collected for five different
persistence
values: $p=0$, $0.25$, $0.5$,
$0.75$, and $1$.  All runs began with an 80\% full lattice, with equal
concentrations of $A$ and $B$ particles, distributed randomly.

In Figure~1, we show a log-log plot of the concentration decay as a function of
time as obtained from simulations.  The data exhibit an asymptotic power-law
decay with a $p$-dependent exponent.  In Figure~2, we plot the local slopes of
the
decay curves as a function of time.  The decay exponents are obtained through
extrapolation to time $t\to\infty$.  Our estimates for the decay exponent for
different values of the persistence parameter are listed in Table~1.

One of the recognized fluctuation effects in two-species reactions is the
tendency of the system to develop large alternating domains of $A$ and $B$
particles [7,13-14,17-18]. We also collected data for some quantities of
interest associated with this phenomenon. These include: \ $\av{l_{AB}(t)}$,
the
average distance between nearest particles of opposite species; \
$\av{l_{AA}(t)}$, the average distance between nearest particles of the same
species; \ $\av{L(t)}$, the average length of a domain of like particles,
measured between the first particles of adjacent domains; and
 \ $\av{n(t)}$, the average number of particles per domain.  In the large-time
asymptotic limit, each of these quantities grows as a power of $t$.
The relation $c\sim\av{n}/\av{L}$ was used to check for consistency. Simulation
results (from about 600 runs on $10^5$-site lattices) are summarized in
Table~1.
The error margins in the table are meant only as rough estimates and may be
overly optimistic.  They merely represent the spread of values we have obtained
by extrapolating the data with different powers of $1/t$.

The case of $p=1$ seems special in that the number of particles per domain
remains bounded and of order unity, even after very long times.  In addition,
the concentration decays as $t^{-1/2}$, similar to isotropic one-species
coalescence, $A+A\to A$. These results are particularly interesting in view of
the following considerations.

Suppose that through fluctuations, large domains of $A$ and $B$ particles were
created, for general $p$. Left alone, each domain moves with the drift
velocity,
while the particle distribution in it is governed by the hard-core constraint
and
is related to the Burgers-equation theory; see [13-14] and references therein.
However, reactions at both ends modify the domain structure. Our results
suggest
that the interplay of the two processes is nontrivial for all $0 \leq p < 1$,
and
the domain size measures involve nontrivial, varying exponents; see Table~1.
Self-organization by formation of large domains has been viewed as an essential
condition for non-mean-field behavior in low-dimensional reactions [6-7,17-18].

However, for $p=1$, we find that there are no large domains. This indicates
that
if a large domain of, for instance, $A$ particles, were formed, it would be
destroyed by $B$ particles ``catching up'' from the left. Indeed, for $p=1$ the
incoming, hopping particle determines the species of the reaction product with
probability
$1$. Thus, regardless of the size of the $A$-domain, its denser, left end will
be
``eaten up'' by $B$ particles faster than its dilute, right end
could catch up with other $A$ particles, while eliminating the $B$ particles
in between. The formation of the shock-like profile due to the interplay of
\UN{hard-core} and \UN{anisotropy} is essential for this phenomenon,
\ie, we need the two domain ends to have different densities.

Once the $p=1$ system is mixed, one would naively anticipate the mean-field
behavior, $\sim 1/t$. However, we know from reactions like $A+A\to 0$, or $A$,
that non-mean-field fluctuations can also arise in ``well-mixed" situations
(no domains), in the form of non-mean-field interparticle distributions [19].
Detailed studies of the
interparticle distribution for the $p=1$ model will be reported in the future.
Since all known one-dimensional reaction-diffusion systems, with or without
hopping isotropy, possess only one  ``diffusive'' length-scale, $\sim t^{1/2}$,
the exponent values of $1/2$ for the $p=1$ model (see Table~1) are probably
exact.

The case of $p=1/2$ resembles the anisotropic two-species
annihilation model,
$A+B\to 0$, studied by Janowsky [12-13] and by Ispolatov \etal\ [14].
Notice that the difference in the number of particles between
the two species, $N_A-N_B$, is locally conserved (in every reaction event) in
the Janowsky [12] model, but only \UN{globally} conserved in the
present model.  This distinction is of no serious consequence in the isotropic
case of no drift [20]. In the case of drift, we argue that the two
models should be similar.  In particular, the heuristic arguments of
Ispolatov \etal\ [14]
could be repeated for our model with $p=1/2$, almost
without change, obtaining the same conclusions and predictions.  Indeed, our
simulation results for
$p=1/2$ are not inconsistent with those already published for $A+B\to 0$.

An interesting observation is that for $p=1/2$ the convergence to the long time
asymptotic behavior of our model seems faster than that of the Janowsky model.
An example is shown in Figure~3, where we plot the time-dependent exponent of
the concentration decay (obtained from local slopes of $\log(c)$ vs.
$\log(t)$), for both cases.  We have no explanation for this phenomenon.
Perhaps it could be exploited to settle the slight discrepancies
between the numerical findings of Janowsky [12-13] and the results of Ispolatov
\etal\ [14].

%
In summary, we have proposed a one-dimensional DLR model with continuously
varying  exponents. It is important to emphasize that our density-exponent
values
were determined largely by numerical studies, and furthermore, the actual
estimates range from $1/2$ for $p=1$ down to near $1/3$ at $p=1/2$, and to
about
$1/4$ for $p=0$. Since all three values: $1/4$, $1/3$, $1/2$, and only these
values, have been encountered for various universality classes of other
one-dimensional reactions, one could suspect that our observation of continuous
exponents is an artefact of finite-time numerics. However, our numerical
simulations were really ``large-scale'' by modern standards, and the data
presented seem to suggest error-limits which clearly favor continuous variation
rather than a smooth crossover between three universal values.

\bigskip
\noindent{\bf Acknowledgments}
\medskip
\noindent We thank S. Redner for numerous useful discussions, and for sharing
his data with us prior to publication.

\vfil\eject
\centerline{\bf References}
\medskip

{\frenchspacing

\item{1.} K.J.~Laidler, {\sl Chemical
Kinetics\/} (McGraw-Hill, New York, 1965).

\item{2.} S.W.~Benson, {\sl The
Foundations of Chemical Kinetics\/} (McGraw-Hill, New York, 1960).

\item{3.} N.G.~van~Kampen, {\sl Stochastic
Processes in Physics and Chemistry\/} (North-Holland, Amsterdam, 1981).

\item{4.} H.~Haken, {\sl Synergetics\/} (Springer-Verlag, Berlin, 1978).

\item{5.} G.~Nicolis and I.~Prigogine, {\sl Self-Organization in
Non-Equi\-librium Systems\/} (Wiley, New
York, 1980).

\item{6.} T.M.~Liggett, {\sl Interacting
Particle Systems\/} (Springer-Verlag, New York, 1985).

\item{7.} K.~Kang and S.~Redner,
{\sl Phys. Rev.} A {\bf 32}, 435 (1985).

\item{8.} V.~Kuzovkov and E.~Kotomin,
{\sl Rep. Prog. Phys.} {\bf 51}, 1479 (1988).

\item{9.} {\sl J. Stat. Phys.} {\bf 65},
nos. 5/6 (1991), Proceedings of {\sl Models of Non-Classical Reaction Rates},
NIH (March 25-27, 1991).

\item{10.} V. Privman, {\sl Dynamics of Nonequilibrium
Processes: Surface Adsorption, Reaction-Diffusion Kinetics, Ordering
and Phase Separation}, in {\sl Trends in Statistical
Physics\/} (Council for Scientific Information,
Trivandrum, India, 1995), in press.

\item{11.} V. Privman, {\sl J. Stat. Phys.} {\bf 72}, 845
(1993).

\item{12.} S. A. Janowsky, {\sl Phys. Rev. E} {\bf 51}, 1858 (1995).

\item{13.} S. A. Janowsky, {\sl Spatial Organization in the Reaction $A+B\to$
inert for Particles with a Drift}, preprint.

\item{14.} I. Ispolatov, P. L. Krapivsky, and S. Redner, {\sl Kinetics of
$A+B\to 0$ with Driven Diffusive Motion}, preprint.

\item{15.} V. Privman, E. Burgos, and M. Grynberg, {\sl Multiparticle Reactions
with Spatial Anisotropy}, {\sl Phys. Rev. E}, in press.

\item{16.} V. Privman, A. M. R. Cadilhe, and M. L. Glasser, {\sl Exact
Solutions of Anisotropic Diffusion-Limited REactions with Coagulation and
Annihilation}, {\sl J. Stat. Phys.}, in press.

\item{17.} D. Toussaint and F. Wilczek, {\sl J. Chem. Phys.}
{\bf 78}, 2642 (1983).

\item{18.} K. Kang and S. Redner, {\sl Phys. Rev. Lett.} {\bf
52}, 955 (1984).

\item{19.} D. ben-Avraham, M. A. Burschka, and C. R. Doering,
{\sl J. Stat. Phys.} {\bf 60}, 695 (1990).

\item{20.} D. ben-Avraham, {\sl Phil. Mag. B} {\bf 56}, 1015 (1987).
}

\vfill\eject
\noindent
\centerline{\bf CAPTIONS}
\bigskip
\bigskip
\bigskip

\NI\hang{\bf Figure 1:} Concentration decay for different values of the
persistence
parameter.  The solid curves represent simulation data for $p=0$, $0.25$,
$0.5$, $0.75$, and $1$ (top to bottom).
\bigskip

\NI\hang{\bf Figure 2:} Local power-law decay exponents for $p=0$, $0.25$,
$0.5$, $0.75$, and $1$ (top to bottom) as a function of inverse time.
The power-law decay exponent are obtained from the local
slopes of the curves in Figure~1.
\bigskip

\NI\hang{\bf Figure 3:} Comparison of the local slopes of $\log(c)$
vs.\ $\log(t)$ between
our model with $p=1/2$ ($*$), and the
anisotropic annihilation model, $A+B\to 0$ ($\circ$).

\vfill\eject

\centerline{\bf Table 1}
\vskip 0.5in

\settabs 6 \columns
\+ {$p$}& {$\alpha$}& {$\beta$}& {$\gamma$}& {$\delta$}& {$\eta$}&\cr

\+ 0     & $0.266(3)$ &$0.292(3)$ &$0.32(1)$ &$0.662(6)$     &$0.41(1)$ &\cr

\+ 0.25  & $0.289(3)$ &$0.311(3)$ &$0.34(1)$ &$0.660(7)$     &$0.39(1)$ &\cr

\+ 0.5   & $0.320(3)$ &$0.345(3)$ &$0.38(1)$ &$0.619(5)$     &$0.309(6)$ &\cr

\+ 0.75  & $0.380(2)$ &$0.400(4)$ &$0.44(1)$ &$0.585(5)$     &$0.215(7)$ &\cr

\+ 1     & $0.499(1)$ &$0.504(6)$ &$0.500(6)$ &$0.499(6)$    &$0.003(4)$ &\cr

\bigskip
\NI\hang{\bf Table 1:} Characteristic exponents for various values of the
persistence parameter,
$p$. The exponents are: $c\sim t^{-\alpha}$, $\av{l_{AA}}\sim t^{\beta}$,
$\av{l_{AB}}\sim t^{\gamma}$, $\av{L}\sim t^{\delta}$, $\av{n}\sim t^{\eta}$.
The error margins represent the spread of values obtained from extrapolations
with various powers of~$1/t$.



\input epsf
\epsfverbosetrue
\epsfxsize 300pt
\epsfysize 200pt
\epsfbox{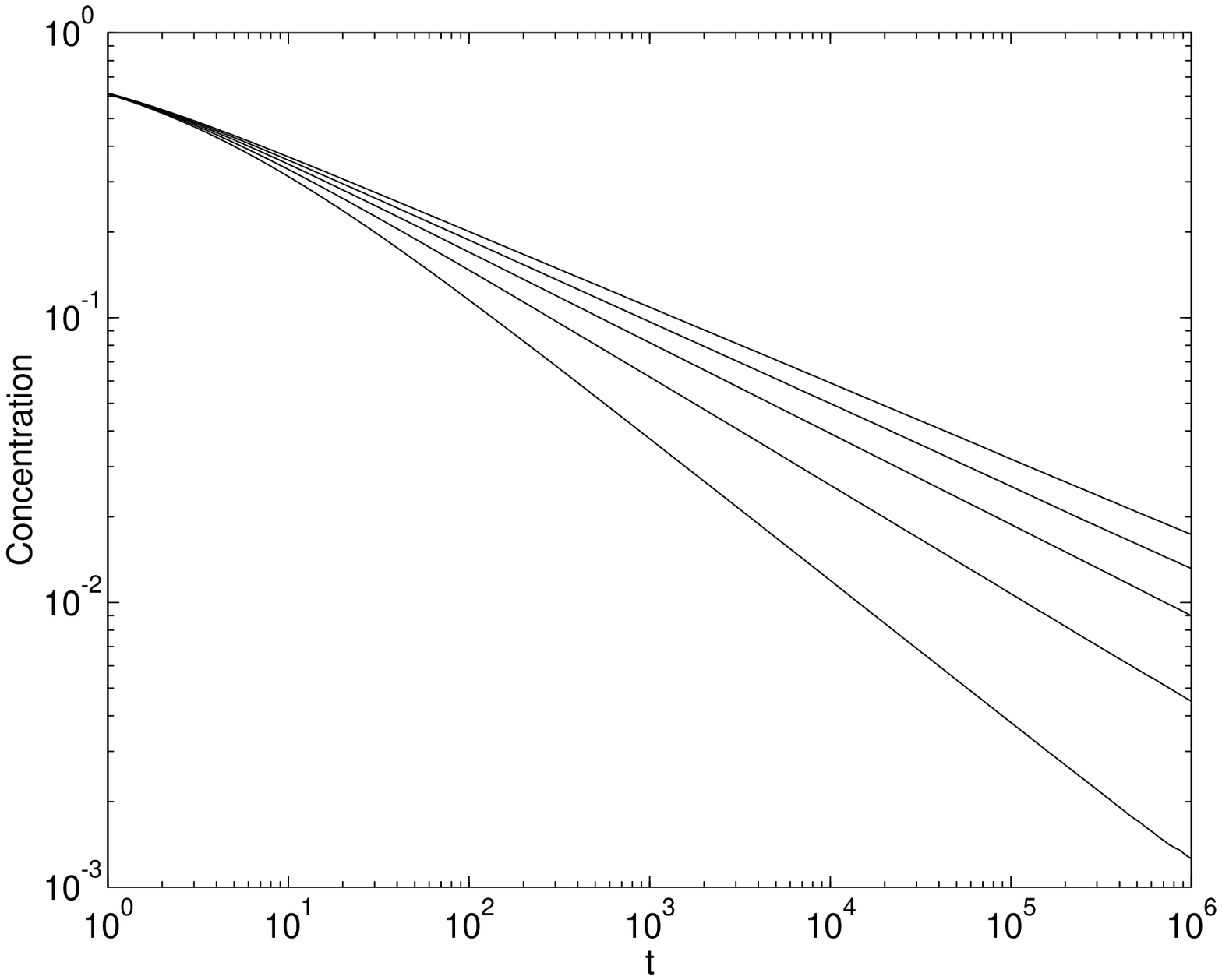}
\epsfxsize 300pt
\epsfysize 200pt
\epsfbox{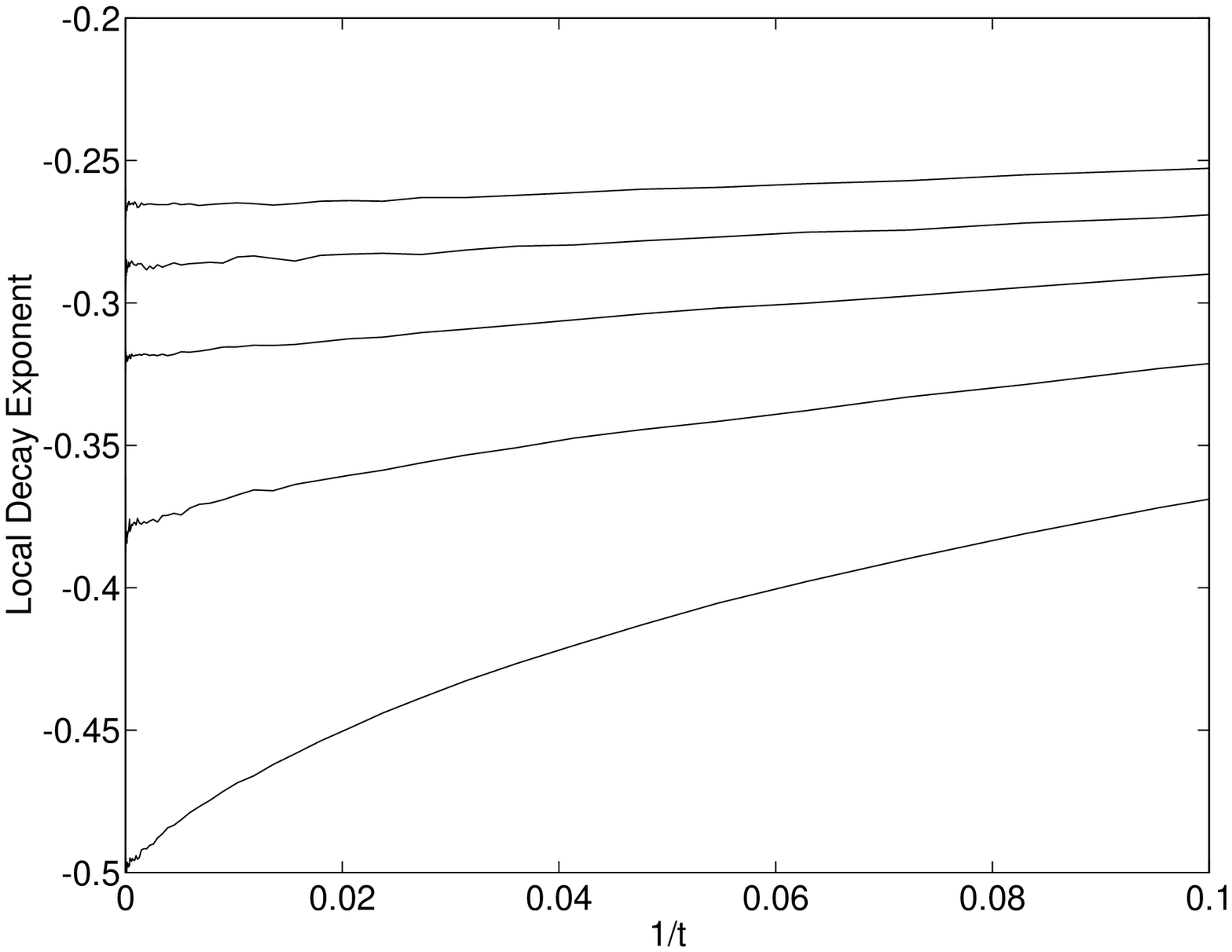}
\epsfxsize 300pt
\epsfysize 200pt
\epsfbox{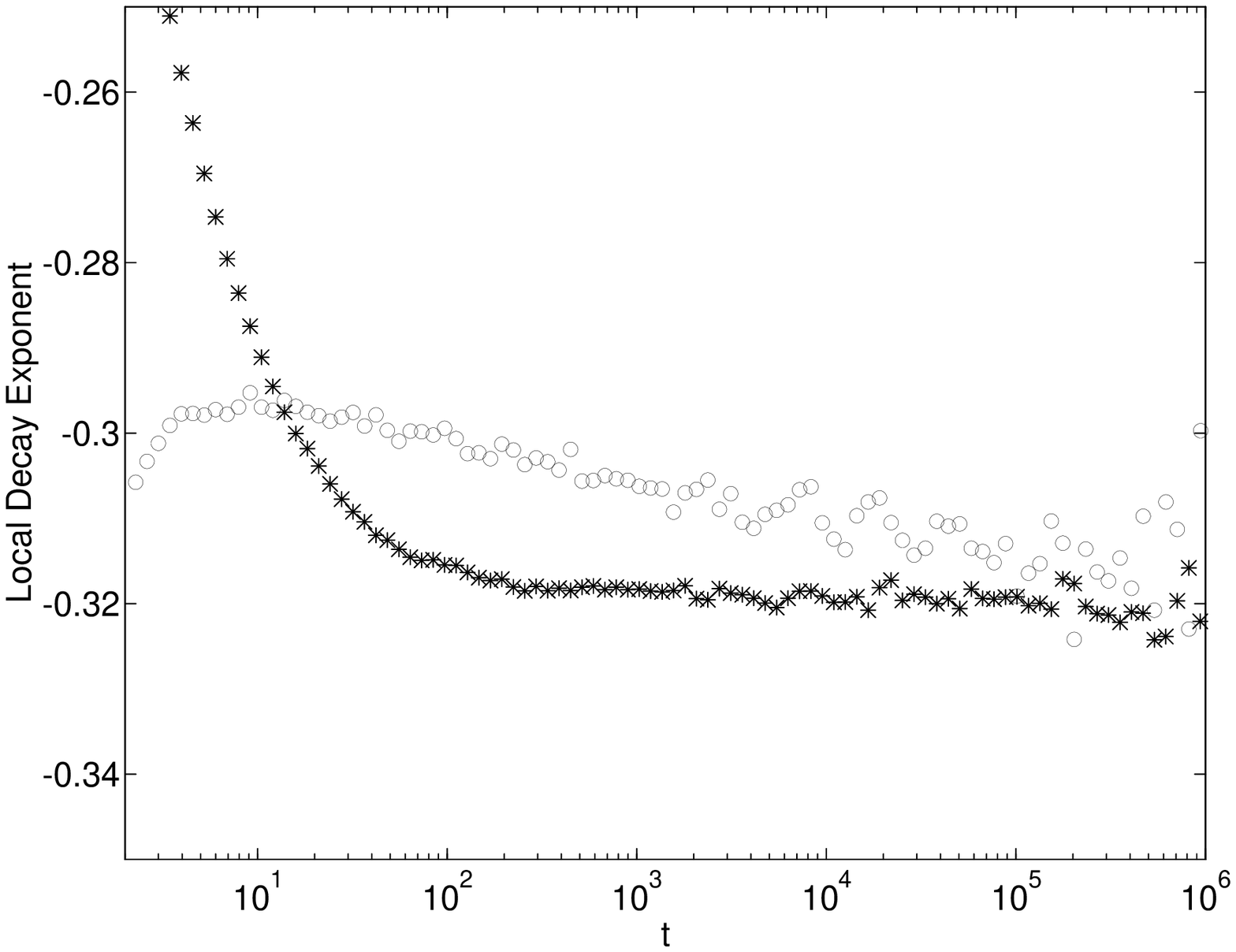}

\bye